\documentstyle[12pt,aaspp4]{article}

\begin{document}

\title{Cluster-Galaxy Correlations in CDM Models}
\author{Manuel E. Merch\'an \altaffilmark{3},
 Mario G. Abadi,\altaffilmark{1}
 Diego G. Lambas  \altaffilmark{2}
and Carlos Valotto \altaffilmark{3}}
\affil{Grupo de Investigaciones en Astronom\'{\i}a Te\'orica y Experimental
(IATE)}
\affil{Observatorio Astron\'omico de C\'ordoba, Laprida 854, 5000, C\'ordoba,
Argentina\\
manuel@oac.uncor.edu, mario@oac.uncor.edu, dgl@oac.uncor.edu,
val@oac.uncor.edu}

\altaffiltext{1}{On a fellowship from CONICET, Argentina}
\altaffiltext{2}{CONICET, Argentina}
\altaffiltext{3}{On a fellowship from CONICOR, C\'ordoba, Argentina}

\begin{abstract}

We study the ability of COBE-normalized CDM models to reproduce 
observed properties of the distribution of galaxies and clusters using N-body 
numerical simulations.
We analyze the galaxy-galaxy and 
cluster-galaxy two-point correlation functions, $\xi_{gg}$ and  $\xi_{cg}$, 
in open ($\Omega_{0}=0.4, 
\Omega_{\Lambda}=0, \sigma_8=0.75$), and flat ($\Omega_{0}=0.3, \Omega_{\Lambda}=0.7,
\sigma_8=1.05$) CDM models
which both reproduce the observed abundances of rich clusters of 
galaxies.

To compare models with observations we compute 
projected cross-correlation functions $\omega_{gg}$ and $\omega_{cg}$
to derive the corresponding $\xi_{gg}$ and $\xi_{cg}$.
We use target galaxies selected from Las Campanas Redshift Survey, target 
clusters selected from the APM Cluster Survey and tracer galaxies from the 
Edinburgh Durham Sky Survey catalog.

We find that the open model 
is able to reproduce the observed
$\xi_{gg}$, whereas the flat model needs antibias in order to fit
the observations.
Our estimate of $\xi_{cg}$ for the APM cluster sample analyzed is consistent
with a power-law 
$\xi_{cg}=({r \over r_{0}}) ^{\gamma}$ with $r_{0}$ = 10.0 $\pm$ 0.7 $h^{-1}$ Mpc
and $\gamma \simeq -2.1$.
For the open and flat-antibiased CDM models
explored we find the corresponding cluster-galaxy correlation lengths $6.5\pm$
0.7 $h^{-1}$ Mpc and 7.2 $\pm$ 0.5  $h^{-1}$ Mpc respectively,
significantly 
lower than the observed value.
Our results indicate that COBE-normalized CDM models are not able to
reproduce the spatial cross-correlation of clusters and galaxies.

\end{abstract}

\keywords{Cosmology-CDM-simulations-correlation function}

\section{INTRODUCTION}

The inflationary scenario and the Cold Dark Matter (CDM) models have become 
one of the most popular theoretical starting point to describe the formation and
evolution of structures in the universe using numerical simulations. Given the
failure of the Standard CDM model (dimensionless density parameter $\Omega_0$=1 and
a Hubble constant $H_{0}=100 h$
 km s$^{-1}$ Mpc$^{-1}$ with $h=0.5$ )
 to reproduce the observed
distribution of galaxies at large scales, several attempts have been made
in order to construct new consistent models.   
The introduction of a cosmological constant ($\Omega_{\Lambda}=
\Lambda/(3 H_{0}^{2}) $) 
in the CDM scenario 
allows for a flat universe ($\Omega=\Omega_0+\Omega_{\Lambda}=1$)
 with $\Omega_0 < 1$ as 
suggested by observations. 
On the other hand, 
measurements of the Cosmic Background Explorer (COBE) satellite have
determined the normalization of 
different power-spectrums of primordial density fluctuations
and therefore 
the present value of the root mean square mass fluctuation 
$\sigma_{8}$ in spheres of radius $8~h^{-1} $ Mpc in the models.
Recently Cole et al. (1997) have analyzed the galaxy-galaxy two-point
correlation function in COBE-normalized CDM models  
with different  
density parameters $\Omega_{0}$ and cosmological constant $\Omega_{\Lambda}$
using numerical simulations.
The authors explore the ability of these CDM models 
to reproduce
observed cluster number densities. Their 
results suggest that COBE-normalized CDM models with 
parameters
$\Omega_{0}=0.4$ , $\Omega_{\Lambda}=0$ and 
$\Omega_{0}=0.3$ , $\Omega_{\Lambda}=0.7$ (with age of the universe
$t_0 \simeq$ 12 and 14 Gyr respectively)
provide a suitable fit to observations.
These models successfully reproduce observed cluster abundances 
 without requiring a strong bias of the distribution of particles in the simulations
 in order to fit the observed galaxy-galaxy correlation function.

The two-point correlation functions are powerful statistical tools to compare 
the observed distribution of galaxies and galaxy clusters and the corresponding model
predictions. The autocorrelation function of bright optically selected 
galaxies is well described by
a power-law fit of the form $\xi_{gg}(r)=(r/r_0)^{\gamma}$ with 
$\gamma=-1.77$ and $r_0=5.4~h^{-1}$ Mpc (\cite{peebles93}, and
references therein).
The joint distribution of galaxies and clusters of galaxies can also be  
statistically described using the cluster-galaxy two-point cross correlation
function 
$\xi_{cg}(r)$.
Seldner \& Peebles (1977)  in their cross-correlation analysis of 
Abell clusters and Lick counts
find a suitable power-law fit $\xi_{cg}(r)=(r/r_0)^{\gamma}$ where $
\gamma \simeq -2$ and $r_{0} \simeq$ 15 $h^{-1}$Mpc. Using similar data,
Lilje \& Efstathiou (1988) 
argue for a lower value of amplitude $r_{0} \simeq 8.8~ h^{-1}$ Mpc
with slope $\gamma \simeq -2.2$. 
The reasons for the different reported amplitudes rely mainly on the assumed
distribution of redshifts of Lick galaxies and deserve further analysis.
Moreover, since
several authors have found dependences of the galaxy-galaxy and
cluster-galaxy correlation lengths
on galaxy luminosity, cluster richness, intracluster gas temperature and velocity 
dispersion,
(\cite{valottob}, \cite{lov95}, \cite{croft97}, \cite{valottoa}) 
 a careful analysis of the target properties is
required to confront properly models and observations.  

In this paper we analyze the distribution of galaxies and clusters of galaxies in 
two COBE-normalized CDM models (open and flat) through numerical simulations. 
We confront the
results of the simulations to observations using 
a sample
of clusters of galaxies taken from the APM cluster catalog and a sample
of galaxies from Las Campanas Redshift Survey (\cite{schectman}).
Section 2 describes the numerical simulations performed and 
section 3 deals with the analysis of the data. In section 4 we confront model
results to observations and 
we analyze the ability of the models to reproduce the observed correlation functions.

\section{NUMERICAL MODELS}

COBE temperature fluctuation measurements 
allows to determine the normalization of 
of the CDM mass fluctuation spectrum for different values of $\Omega_{0}$ and
$\Omega_{\Lambda}$. 
In Figure 1 are plotted 
$\Omega_{0}$ as a 
function of $\sigma_{8}$ 
in COBE-normalized CDM models extracted from Table 1 of G\'orski et al.
(1995) and Cole et al. (1997).
The solid and dashed thin lines 
correspond to open models with $\Omega_{\Lambda}=0$ 
and flat models with $\Omega_{\Lambda} \ne 0$, respectively.
In this figure it is also shown (thick line) the corresponding relation 
between these parameters 
found in open CDM models (dashed) and flat (solid)  
corresponding to the fit of the cluster temperature function 
computed by Eke, Cole \& Frenk (1996).
The intersection of these curves provide the suitable values 
of the parameters that fit simultaneously cluster abundances and COBE
normalizations.
By inspection to this figure it is apparent our choice of models: 
open, with $\Omega=0.4, ~\Omega_{\Lambda}=0, ~\sigma_8=0.75$
; and flat, with $\Omega=0.3, ~\Omega_{\Lambda}=0.7, \sigma_8=1.05$ (both with a Hubble parameter $h=0.65$)
that fulfill this condition.
The vertical lines show the allowed range of
values of $\sigma_{8}$ compatible with the observed
relative fluctuations in the number of bright galaxies $\delta N/N=1.35(r/r_0)^
{\gamma/2}$, where $\gamma=-1.77\pm 0.04$ and $r_0=5.4\pm 1.0 h^{-1}$Mpc
(\cite{peebles93}).
Thus, no strong biasing of the spatial distribution of particles is required to 
infer the properties of the galaxy distribution in these models.

For these two models (open and flat) we have performed N-body numerical simulations using the Adaptative
Particle-Particle Particle Mesh (AP3M) code
developed by Couchman (1991).
Initial positions and velocities of particles were generated using the Zeldovich 
approximation  
corresponding to the CDM power spectrum. 
The computational volume is a periodic cube of side length 195 $h^{-1}$ Mpc.
We have followed the evolution of $N=5\times 10^{5}$ particles
 in a $64^{3}$ grid
mesh and 4 levels of refinements as a maximum. 
The resulting mass per particle is $4.11 \times 10^{12} 
\Omega_{0} h^{-1} M_{\sun} $.
We have adopted an
analytic fit to the CDM power spectrum of the form 

\begin{equation}
p(k) \propto \frac{k}{[1+3.89q+(16.1q)^2+(5.46q)^3+(6.71q)^4]^{1/2}}
\left( \frac{ln(1+2.34q)}{2.34/q} \right) ^{2}
\end{equation}

where $q=(k/\Gamma) h$ Mpc $^{-1}$, $\Gamma=\Omega_{0}~h$ exp-$(\Omega_B+
\Omega_B\Omega_0)$ and $\Omega_B=0.0125~h^{-2}$ is the value of the baryon density
parameter imposed by nucleosynthesis theory (\cite{bardeen} and \cite{sugiyama}).
The initial conditions correspond to redshift $z=10$ and
the evolution was followed using 1000 steps. 

We identify centers of mass of clumps of particles in the simulations
using a standard friends-of-friends algorithm with a
linking length $l=0.17~n^{-1/3}= 
418h^{-1}$ kpc, 
where $n$ is mean particle density
(\cite{cole}). 
Using these centers we define the clusters in the simulations as the    
particles within Abell radius $R_{A}= 1.5h^{-1}$Mpc and compute
the corresponding cluster masses.   
In figure 2 we show the resulting cumulative mass function of 
the two CDM models at redshift $z=0$ and the analytic fit to
observations given by Bahcall \& Cen (1993). Following Cole et al.
(1997) we show a box indicating the mass range of clusters with 
observed abundance $4 \times 10^{-6} ~h^{3}$ Mpc $^{-3}$.
As it can be appreciated in this figure there is a good agreement
between observations and the two models analyzed consistent 
with Cole et al. (1997) results. 
 
\section{ANALYSIS OF OBSERVATIONS}

In this section, we compute the projected two-point cluster-galaxy
 and galaxy-galaxy cross-correlation function ($\omega_{cg}(r_p)$ and $\omega_{gg}(r_p)$, 
respectively), where $r_p$ is the projected distance between targets 
(clusters or galaxies) and tracers (galaxies). 
We estimate the projected target-tracer cross-correlation
function using 
\begin{equation}
\omega(r_p)=\frac{\langle N(r_p) \rangle }{\langle N_{RAN}(r_p)\rangle }-1, 
\end{equation}
where $\langle N(r_p) \rangle$ is the mean  
number of target-tracer pairs separated by a projected  
distance $r_{p}$ in the data and 
$\langle N_{RAN}(r_p) \rangle $ corresponds to targets with random
angular positions and with the same redshift distribution than the data targets.

The determination of the spatial correlation function $\xi(r)$ from $\omega(r_p)$
requires the inversion of  

\begin{equation}
w(r_{p})= C \int_{-\infty}^{\infty} \xi((\Delta ^{2}+r_p^{2})^{1/2})
~ d \Delta 
\label{wr}
\end{equation}
This integral extends over all line-of-sight separations $\Delta $ of target-tracer pairs.
The constant C in equation~(\ref{wr}) is related to the probability that a tracer 
galaxy is found at a radial distance $d$ from the observer.
Assuming a power-law model for the 
cross-correlation function $\xi(r)=(r/r_0)^{\gamma}$ Lilje \& Efstathiou
(1988) derive

\begin{equation}
w(r_{p})=C \sqrt{\pi}\frac{\Gamma[-(\gamma +1)/2]}{ \Gamma(-\gamma/2)} \frac
{ r_0^{-\gamma}}{ r_p^{-(\gamma+1)}}
\label{wr0}
\end{equation}

Using equation (2) we have computed $\omega(r_p)$, and by fitting power-laws we have inferred
correlation lengths $r_0$ and slopes $\gamma$ from equation (4).  

We have chosen tracer objects corresponding to galaxies in the 
southern galactic hemisphere 
of the Edinburgh-Durham sky survey (hereafter COSMOS survey).
Angular positions and apparent $B_j$ magnitudes are available for all galaxies 
in COSMOS.
At faint magnitudes $B_j$ mis-classification of stars and galaxies, 
plate zero-points and photometric errors 
become critical. Taking this fact into account, and     
in order to check our estimates of the correlation 
function fitting
parameters we have defined two samples of tracers 
with limiting $B_j$ magnitudes 
$m_{lim} = 18.0$ and $m_{lim} = 19.0$. 
We have selected target clusters 
from the APM Cluster Survey (\cite{dalton94}) restricting 
our analysis to clusters with
APM richness 30 $<$ $\cal R$ $<$ 60 and  with 
radial velocities in the range 10,000-40,000 km s$^{-1}$ since the number density of clusters 
falls rapidly beyond 40,000 km s$^{-1}$. 
The lower limit in radial velocity was adopted in order to avoid 
large angular separations in the computation of correlations.
The restriction on APM richness $\cal R$ is based on the fact that clusters with
 $\cal R$ $ <$ 30
are very poor and their number density continuously fall beyond 15000 km/s.
There are only 8 clusters with $\cal R$  $>$ 60, these objects were 
also not included in our studies
since their radial velocities are beyond the mean of our sample ($\simeq$ 25000 km/s).   
It should also be remarked that the selection procedure used to build 
the APM Cluster Survey makes it free from artificial inhomogeneities 
(\cite{dalton97}).  
We have considered two subsamples according to the richness parameter:
30 $<$ $\cal R$ $<$40 and 40 $<$ $\cal R$ $<$ 60 in order to search for possible
 richness dependences of cluster-galaxy correlations.
To check the consistency of our results we have also computed $\xi_{cg}$ 
using samples of clusters taken from David et al. (1993) and Ebeling et
al. (1996)
which provide intracluster temperatures; and from Fadda et al. (1996) which 
provide estimates of cluster velocity dispersions.
From these samples we selected clusters with radial velocities in the same 
range than that adopted for the APM clusters and the corresponding analysis 
serves as an independent reproducibility test of our results.

The sample of target galaxies is taken from Las Campanas Redshift Survey 
(hereafter LCRS), Schectman et al (1996).
The average radial velocity of these galaxies is $\simeq$ 30,000 km s$^{-1}$ 
and extends to $\simeq$ 80,000 km s$^{-1}$. 
Given uncertainties in the derivation of spatial correlations from
projected data we have attempted to focus our analysis on targets with similar
redshift distributions since real differences of spatial correlations
among the samples would directly reflect in the projected correlation functions. 
The distribution function of radial velocities of LCRS galaxies is shown in solid line
in figure 3. 
For comparison is
also shown with dashed line the corresponding distribution of 30 $<$ $\cal R$ $<$ 60
APM clusters. 
The dotted line in the figure indicates the resulting  
distribution of LCRS galaxies radial velocities 
where a radial gradient is imposed through a Monte-Carlo rejection
algorithm that gives a similar 
distribution than the APM cluster sample
, hereafter restricted LCRS galaxies. 

We have adopted power-law fits $\omega(r_p)=A r_{p}^{(\gamma+1)}$ in the range
0.2 $h^{-1}$Mpc $<r_{p}<$ 10 $h^{-1}$Mpc and  
0.2 $h^{-1}$Mpc $<r_{p}<$  4 $h^{-1}$Mpc   
for cluster-galaxy and galaxy-galaxy correlations respectively. 
In figure 4 we show $\omega_{gg}(r_p)$ and $\omega_{cg}(r_p)$ and the corresponding 
power-law least squares fits. Error bars correspond to estimates derived from
the bootstrap resampling technique developed by Barrow, Bhavsar \&
Sonoda (1984) with 30 bootstrap target samples. 
For the derivation of $r_0$ from equation (4) it is necessary to estimate the 
constant $C$
as an integral that includes the luminosity function in the case of a magnitude limited sample
of tracers
such as COSMOS catalog. For this purpose,
we use a Schechter function fit to the luminosity function of COSMOS galaxies
with parameters $M^*=-19.50 \pm 0.13, \alpha=-0.97 \pm 0.15$ (\cite{lov92}), 
a K-correction term of the form $3z$, and a flat cosmology ($\Omega_{0}=1$). 
It should be recalled the various sources of
error involved in the determinations of the values of $r_{0}$ through the inversion of 
equation (\ref{wr})
such as uncertainties in the luminosity function parameters,
K-corrections and cosmological model, as well as observational biases involving
selection effects, photometric errors, etc.
The results of our statistical analysis are shown in table 1. 
The quoted errors in the values of $r_{0}$ were derived through propagation from
the rms errors in the $\omega_{gg}$ and $\omega_{cg}$ power-law fits and
variations in C due to uncertainties in the luminosity function parameters, added in quadrature.

From
inspection to this table one can notice that the value of the correlation length
of the galaxy-galaxy correlation function for the restricted LCRS sample
is $r_{0gg} = 3.8 \pm 0.4 ~h^{-1}$Mpc, lower than
the standard value of 5.4 $h^{-1}$Mpc. 
This is mostly probably due to a luminosity effect 
(\cite{lov95}, \cite{valottob})
given that
the majority of target galaxies in this sample are $L<L^{*}$. 
The values of the cluster-galaxy 
cross-correlation lengths shown in table 1
are $\simeq 10-30 \% $ larger than those derived by Lilje \& Efstathiou
(1988) (except for the $30 < $ $\cal R$ $<40$ APM cluster sample)
where it may be argued that these differences may rely on the
selection function adopted for the Lick catalog. It can also be seen
in this table that the richest clusters have larger values of $r_{0cg}$
(\cite{valottoa}). 
For comparison we have also computed the cross-correlation function for a
sample of Abell clusters with measured temperatures and velocity dispersions
(\cite{ebeling}, \cite{fada}) in the same range of radial velocities.
These samples, although with a small number of targets also show somewhat 
large values of $r_{0cg}$ compared to Lilje \& Efstathiou (1988) results, 
consistent with our estimates of APM clusters and giving additional support to our
results.
It should be remarked that the derived values of the
cluster-galaxy correlation length found are not likely to be biased high due
to systematics or deprojection calculations given the relatively low value
of the galaxy-galaxy correlation length obtained for targets with an equivalent redshift distribution.

\section{COMPARISON BETWEEN MODELS AND OBSERVATIONS}

In order to compare the results of observations and numerical models 
we have used 30 $<$ $\cal R$ $<$ 60 APM clusters taking advantage of 
the statistically significant number of target objects in a well defined volume 
and the fact that the APM cluster catalog is free from projection and subjective biases.
In order to make an appropriate comparison with observations we have
attempted to select a 
subsample of clusters in the numerical simulations with comparable
characteristics
to this APM cluster sample. Since the APM cluster catalog provides
a suitable richness parameter $\cal R$  
we have used 18 APM clusters with measured line of sight velocity dispersions $\sigma$ 
(\cite {fada}) to provide a suitable relationship between $\sigma$
and   
APM cluster richness $\cal R$.  
We have also added 15 APM clusters with measured temperatures $T$ (\cite{ebeling})
using $\sigma=400  T^{1/2}$ where $\sigma$ is km/s and $T$ in KeV.
Given the dispersion of the correlation
between $\cal R$ and $\sigma$ a simple
linear relation of the form 
$\cal R$= $\sigma /19 +10 \pm 10$ provides a suitable fit to the data.  
We assign an equivalent APM cluster richness
$\cal R$ to the clusters in the simulations applying 
this relation to 
the actual radial velocity dispersions of the simulated clusters
and select a set of clusters with the same $\cal R$ distribution 
and number density (10$^{-5}$ h$^{3}$ Mpc$^{-3}$) than our $30<$ $\cal R$ $<60$ APM cluster sample.
This procedure provides a suitable
set of clusters in the numerical simulations that can be confronted
to observations. 

First, we have considered each particle a galaxy in both
open and flat CDM numerical simulations.
We have computed the 
cluster-galaxy $\xi_{cg}(r)=\langle N_{cg}(r)\rangle / \langle N_{RAN}(r)\rangle -1$ and galaxy-galaxy 
$\xi_{gg}(r)=\langle N_{gg}(r)\rangle /\langle N_{RAN}(r)\rangle -1$ two-point 
cross-correlation functions in the
numerical simulations where $\langle N_{cg}(r) \rangle$, 
$\langle N_{gg}(r) \rangle$ and $\langle N_{RAN}(r) \rangle$ are the
mean number of cluster-galaxy, galaxy-galaxy and random-galaxy 
pairs at spatial separation $r$. 
The derived cluster-galaxy correlation functions
can be fitted by power-laws  
in the range 2 and  20 $h^{-1}$Mpc.
In table 2 
are listed the corresponding values
of correlation length $r_{0}$ and slope $\gamma$ of
the power-law fits to
 $\xi_{cg}(r)$ and $\xi_{gg}(r)$ in the models. 
Quoted uncertainties in this table correspond to errors of the least squares fits.
Errors in $r_{0cg}$ have added in quadrature the corresponding dispersion 
of values due to the spread in the observed $\cal R$-$\sigma$ relation. 
By comparison of tables 1 and 2, and in agreement with Cole et al.
(1997) it can be seen that the open model requires practically no bias, 
while the flat model a moderate anti-bias.
Due to the failure of the flat model to reproduce the observations
we have generated different mock catalogs of 'galaxies' for
this model associating a probability $P$ of the particles being a galaxy 
according
to different prescriptions.
We smooth the density field calculating
the density $\eta$ centered in each particle within a sphere of radius
 $1.5 h^{-1}$  Mpc. We have adopted two different models for $P(\eta)$:
a power law, $P(\eta) = (\eta /\eta_{c})^{\alpha}$ and
a step function $P(\eta) = 0$ if $\eta > \eta_{min}$, $P(\eta)=1$ otherwise.
Both models are constrained to provide a $\xi_{gg}$ consistent 
with observations. 
We have found that the cluster-galaxy cross-correlation function
$\xi{cg}$ of the models for different parameters are very similar. 
This is expected
since the antibias is not too strong and the imposed observational
constrain on $\xi_{gg}$.
The resulting power-law fitting parameters of $\xi_{gg}(r)$ with
$\alpha =1$ and a suitable value of $\eta_{c}$  are
shown in table 2 where it can be seen the good agreement 
with observations for this simple power-law anti-biasing model. 
Nevertheless 
$\xi_{cg}$ of the models do not
fit the observations,  
the cluster-galaxy cross-correlation lengths of the open and
flat antibiased CDM models are $\simeq 25 \%$ lower than observed
with a statistically significant confidence.
In figure 5 are shown the cluster-galaxy correlation function of the models and the
observations where it is apparent the 
inconsistency of the models. 
In order to check the statistical stability of our results 
we have taken into account the observed dispersion in the 
$\cal R$ -$\sigma$ relation in the selection of clusters in the numerical simulations
. We find negligible changes in the
correlation lengths of the models suggesting that our results are not
too strongly dependent on the particular selection of the simulated clusters. If
only the $\simeq$ 10 richest clusters are used in the cross-correlation 
analysis, significantly
larger values of $r_{0} \simeq 9$ h$^{-1}$ Mpc may be obtained. 
Certainly this cannot be used for 
a proper comparison to observations since the abundance of APM
clusters $\gtrsim 1 \times 10^{-5}$ $h^{3}$ Mpc$^{-3}$ corresponds 
to  $\gtrsim 75$ clusters in our 
computational volume. 
 
The derivation of spatial correlations involve the selection function of COSMOS
galaxies and therefore play an important role uncertainties in the luminosity function parameters,
K-corrections, etc. The ratio of correlation functions in the power law approximation
writes: 

\begin{equation}
\frac{\xi_{cg}(r)}{\xi_{gg}(r)}= \frac{A_{cg}~ f(\gamma_{gg})}{A_{gg}~f(\gamma_{cg})}
r^{(\gamma_{cg} - \gamma_{gg})}
\end{equation}

where $f(\gamma)={\Gamma(-(\gamma+1)/2) \over \Gamma(-\gamma/2)}$, $\Gamma$ is the gamma
function and $A$ and $\gamma$ refer to the amplitude and slope of the projected correlation
fits, defined in section 3.
This ratio is independent of the deprojection uncertainties
already mentioned and therefore provide 
an unbiased measure of the relative clustering of galaxies and galaxy clusters. 
Figure 6 displays the ratio of the cluster-galaxy to the galaxy-galaxy correlation functions
$\xi_{cg}/\xi_{gg}$ 
for the numerical simulations and the observations.
It can be seen in this figure the large disagreement between the
observations and the model results  
showing that observed clusters 
are in higher density galaxy 
environment than the simulated clusters in the CDM models explored.

\section{CONCLUSIONS}

We have tested an open, 
and a flat $\Lambda$ dominated COBE
normalized CDM model 
through the computation of 
the cluster-galaxy cross-correlation
function in order to shed new light on the observational 
viability of this structure formation scenario.
Our analysis of the cluster-galaxy cross-correlation function provides
a significant statistical evidence of the failure of COBE normalized CDM 
models to reproduce
the extended halos of clusters.
In spite of the success of these models in reproducing the abundance 
of rich clusters 
 and the general pattern of galaxy clustering,
  the high observed amplitude of cluster-galaxy correlations cannot
 be reproduced in the models.
These results are consistent with the virial analysis of clusters
in Cole et al. (1997) where a positive bias is needed in the
flat $\Lambda$ dominated CDM model to fit observed cluster M/L ratios.
 
A detailed comparison of models and observations should be stressed since
we have found significant dependences of the cross-correlation function
parameters on the velocity dispersion of the clusters in both simulations
and observations. 
Our results rely on a well controlled sample of galaxy clusters 
as well as on a comparable set of clusters from numerical simulations 
giving confidence on our results against the ability of COBE-normalized CDM 
models to reproduce the joint distribution of galaxies and clusters.

\acknowledgments

We thank H. Couchmann for kindly providing the AP3M code and
Robert Kirshner for kindly permitting the use of the Las Campanas
Redshift Survey. 
This work was partially supported by the Consejo de Investigaciones Cient\'{\i}ficas
y T\'ecnicas de la Rep\'ublica Argentina, CONICET, the Consejo de
Investigaciones Cient\'{\i}ficas y Tecnol\'ogicas de la Provincia de C\'ordoba, CONICOR
and Fundaci\'on Antorchas, Argentina.

 {}
\newpage

\figcaption{Constraints on $\Omega_0$ and $\sigma_8$ in CDM models. 
The thin lines correspond to COBE normalization for the open model 
with $\Omega_{\Lambda}=0$ model (solid), and the flat model with 
$\Omega_{\Lambda} \ne 0$ (dashed). 
Similarly, thick lines represent cluster abundances derived by 
Eke, Cole \& Frenk  1996.
The vertical lines correspond to the rms fluctuation in the number of 
galaxies taking into account the quoted errors in the galaxy-galaxy 
correlation length (see text).  
Horizontal error bars for the open model are taken from table 1 of 
G\'orski et al. 1995
and for the flat model the error bars correspond to 
$1\sigma$ deviations taken 
Cole et al. 1997.
}

\figcaption{Observed cluster abundances given by 
Bahcall \& Cen 1993
(dotted line) and the correponding abundances
inferred from the open (solid line) and flat (dashed line) models. The box 
indicates the observational range of masses with abundance $4 \times 10^{-6}$
h$^3$ Mpc$^{-3}$ }

\figcaption{Distribution of radial velocities of clusters 
and galaxies. $30 <$ $ \cal R$ $ < 60$ APM  clusters, dashed line;
complete LCRS galaxies, solid line; and restricted LCRS galaxies, dotted line. } 
\figcaption{Observed projected cross-correlation
functions. The circles show $\omega_{cg}(r_{p})$ for our sample of 96 
APM 30$ <$ $\cal R$ $< 60 $ cluster targets and COSMOS survey tracer
galaxies
with limiting magnitude $m_{lim}=18$. The solid, long dashed and dotted lines correspond to 
power-law fits 
 of the cluster target samples 30 $<$ $\cal R$ $<$60, 30$ <$ $\cal R$ $< 40 $ 
and 40$ <$ $\cal R$ $< 60 $ respectively.
The triangles show  
$\omega_{gg}(r_{p})$ for the restricted LCRS target galaxies  
and the same COSMOS tracers. The short dashed line shows the corresponding power-law fit.  }

\figcaption{Spatial cluster-galaxy cross-correlation function. 
The dashed line corresponds to the power-law fit 
$\xi_{cg}=({r \over 10 h^{-1} Mpc}) ^{-2.09}$ of the APM cluster sample 
30 $<$ $\cal R$ $<$ 60. 
The circles, squares and triangles correspond to 
  $\xi_{cg}$ of the open, flat and flat antibiased CDM models respectively. }

\figcaption{Ratio of cluster-galaxy to galaxy-galaxy correlation functions
in the numerical models and the observations. 
The smooth solid line indicates the ratio of the power-law fits (eq. 5) corresponding to 
30 $<$ $\cal R$ $<$ 60 APM clusters and the restricted LCRS galaxies
($m_{lim}=18$). Circles, squares and triangles
correspond to the open flat and flat biased CDM models respectively.  }

\newpage

\begin{deluxetable}{ccccccc}
\tablewidth{40pc}
\tablecaption{Observational Results}
\tablehead{
\colhead{ Sample}&
\colhead{ $N$}&
\colhead{ m$_{lim}$ } &
\colhead{ $C ~ [\times 10^{-3}]$}&
\colhead{ $A$}&
\colhead{ $\gamma$ }&
\colhead{ $r_0$ (h$^{-1}$ Mpc)} 
 }
\startdata
Galaxies                   & \nl
LCRS (restricted)          & 3033 & 18 & 3.15 & $0.16 \pm 0.01$ & $-1.91 \pm 0.06$ & $3.8  \pm 0.4$ \nl
LCRS (restricted)          & 3033 & 19 & 2.34 & $0.08 \pm 0.01$ & $-1.91 \pm 0.04$ & $3.5  \pm 0.3$ \nl
Clusters                   & \nl
APM 30 $< $ $\cal R$ $<$60 & 96   & 18 & 3.14 & $1.09 \pm 0.05$ & $-2.09 \pm 0.05$ & $10.0 \pm 0.7$ \nl
APM 30 $< $ $\cal R$ $<$60 & 96   & 19 & 2.41 & $0.96 \pm 0.04$ & $-2.04 \pm 0.04$ & $10.8 \pm 0.6$ \nl
APM 30 $< $ $\cal R$ $<$40 & 64   & 18 & 3.22 & $0.97 \pm 0.07$ & $-2.13 \pm 0.07$ & $8.9  \pm 0.8$ \nl
APM 40 $< $ $\cal R$ $<$60 & 32   & 18 & 2.94 & $1.42 \pm 0.07$ & $-2.05 \pm 0.04$ & $12.0 \pm 0.8$ \nl
Fadda et al clusters       & 18   & 18 & 3.28 & $1.09 \pm 0.08$ & $-2.00 \pm 0.06$ & $10.3 \pm 1.0$ \nl
Ebeling et al clusters     & 18   & 18 & 3.34 & $1.59 \pm 0.08$ & $-2.07 \pm 0.04$ & $11.6 \pm 0.8$ \
\enddata
\end{deluxetable}

\begin{deluxetable}{ccccc}
\tablewidth{35pc}
\tablecaption{Model Results}
\tablehead{
\colhead{ $Model$}&
\colhead{ $\gamma_{gg}$ }&
\colhead{ $r_{0gg}$ (h$^{-1}$ Mpc) }&
\colhead{ $\gamma_{cg}$ }&
\colhead{ $r_{0cg}$ (h$^{-1}$ Mpc)}
 }
\startdata
Open         & $-2.20 \pm 0.08$ & $4.10 \pm 0.41$&$ -2.14 \pm 0.09$ &$ 6.54 \pm 0.68$ \nl
Flat         & $-2.06 \pm 0.05$ & $5.90 \pm 0.41$&$ -1.88 \pm 0.09$ &$ 8.39 \pm 0.83$ \nl
Flat Bias    & $-1.90 \pm 0.07$ & $4.22 \pm 0.39$&$ -1.86 \pm 0.06$ &$ 7.22 \pm 0.47$ \
\enddata
\end{deluxetable}


\begin{thebibliography} {}
\bibitem[Bahcall \& Cen  1993]{bahcall} Bahcall, N.A. \& Cen, R. 1993,
\apj, 407, L49
\bibitem[Bardeen et al.  1986]{bardeen} Bardeen, J.M., Bond, J.R.,
Kaiser, N., \& Szalay, A.S. 1986, \apj, 306, 15
\bibitem[Barrow, Bhavsar and Sonoda  1984]{barrow} Barrow, J.D., Bhavsar,
S.P. \& Sonoda, D.H., \mnras, 210, 19p
\bibitem[Cole et al.  1997]{cole} Cole, S., Weinberg, D.H., Frenk, C.S. \& Ratra, B.
1997, \mnras, in press.
\bibitem[Couchman  1991]{couchman} Couchman, H.M.P. 1991, \apj, 368, L23
\bibitem[Croft et al.  1997]{croft97} Croft, R.A.C., Dalton, G.B., Efstathiou, G. \&
Sutherland, W.J., 1997, astro-ph, 9701040 
\bibitem[Dalton et al.  1994]{dalton94} Dalton G.B., Efstathiou G., Maddox S.J.
\& Sutherland W.J. 1994, \mnras, 269, 151 
\bibitem[Dalton et al.  1997]{dalton97} Dalton, G.B., Maddox, S.J., Sutherland, W.J. \&
Efstathiou, G., 1997, astro-ph, 9701180 
\bibitem[David et al.  1993]{david} David, L.P., Slyz, A., Jones, C.,
 Forman, W., Vritilek D.  1993, \apj, 412, 479 
\bibitem[Eke, Cole \& Frenk   1996]{eke96}Eke, V.R., Cole, S. \& Frenk, C.S., 1996, \mnras, 282,
263
\bibitem[Ebeling et al.  1996]{ebeling} Ebeling, H., Voges, W., B\"{o}hringer, H., Edge, A.C.,
Huchra, J.P. and Briel, U.G., 1996, \mnras, 281, 799
\bibitem[Fadda et al.  1996]{fada} Fadda, D., Girardi, M., Giuricin, G., Mardirosian, F. \& Mezzetti,
M., 1996, \apj, 473, 670
\bibitem[G\'orski et al.  1995]{gorski} G\'orski, K.M., Ratra, B., Sugiyama, N. \& Branday
A.J., 1995, \apj, 444, L65 
\bibitem[Henry \& Arnaud  1991]{henry} Henry, J.P. \& Arnaud, K.A. 1991, \apj, 372, 410 
\bibitem[Lilje \& Efstathiou  1988] {lilje} Lilje, P.B. \& Efstathiou G., 1988, \mnras, 231, 635 
\bibitem[Loveday et al.  1992] {lov92} Loveday, J., Peterson, B.A., Efstathiou, G. \& Maddox,
S.J.,1992, \apj, 390, 338 
\bibitem[Loveday et al.  1995]{lov95} Loveday, J., Maddox, S.J., Efstathiou, G.
\& Peterson, B.A., 1995, \apj, 442, 457
\bibitem[Navarro, Frenk \& White  1995]{navarro} Navarro, J.F., Frenk, C.S. \& White, S.D.M,
1995.\mnras, 275, 720
\bibitem[Peebles   1993]{peebles93}  Peebles, P.J.E., 1993, 
Principles of Physical Cosmology, Princeton University Press, Princeton. 
\bibitem[Schectman et al.  1996]{schectman}Schectman, S., Landy, S., Oelmer, A., Tucker, D., Lin,
H., Kirshner, R., \& Schechter, P. 1996, \apj, 464, 60
\bibitem[Seldner \& Peebles  1977]{seldner} Seldner, M. \& Peebles, P.J.E., 1977, \apj, 215, 703
\bibitem[Struble \& Rood  1991]{struble} Struble M.F. \& Rood, H.J., 1991, \apj, 77, 363
\bibitem[Sugiyama  1995]{sugiyama} Sugiyama, N. \apjs, 100, 281
\bibitem[Smoot et al.  1992] {smoot} Smoot G. 1992, \apj, 396, L1
\bibitem[Valotto \& Lambas  1995]{valottoa} Valotto, C.A. \& Lambas D.G., 1995, \mnras, 277, 896
\bibitem[Valotto \& Lambas  1997]{valottob} Valotto, C.A. \& Lambas D.G., 1997, \apj, 281, 896
\bibitem[White \& Fabian  1995]{whitefabian}  White D.\& Fabian, G, 1995, \mnras, 273, 72
\end{thebibliography}
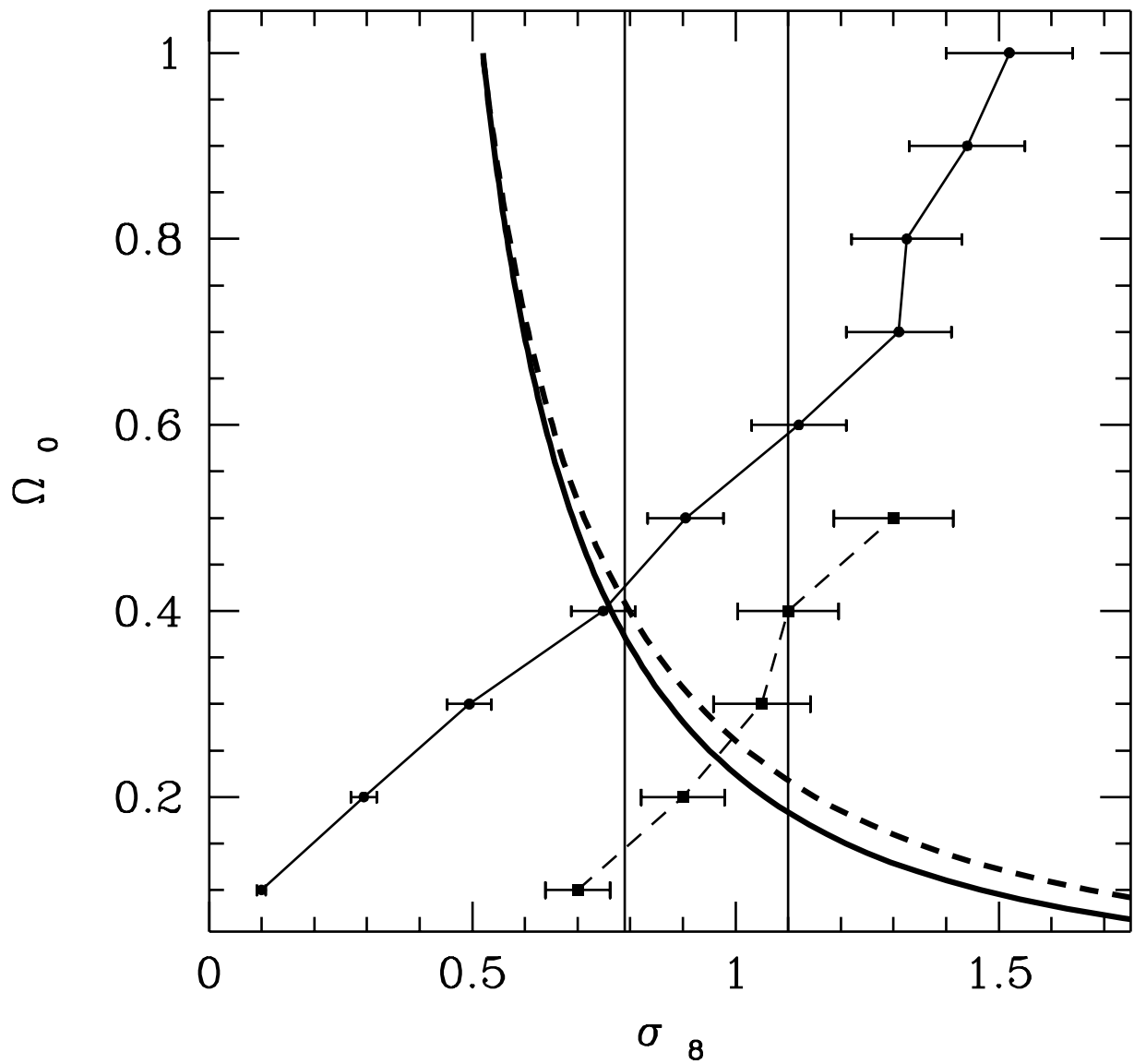
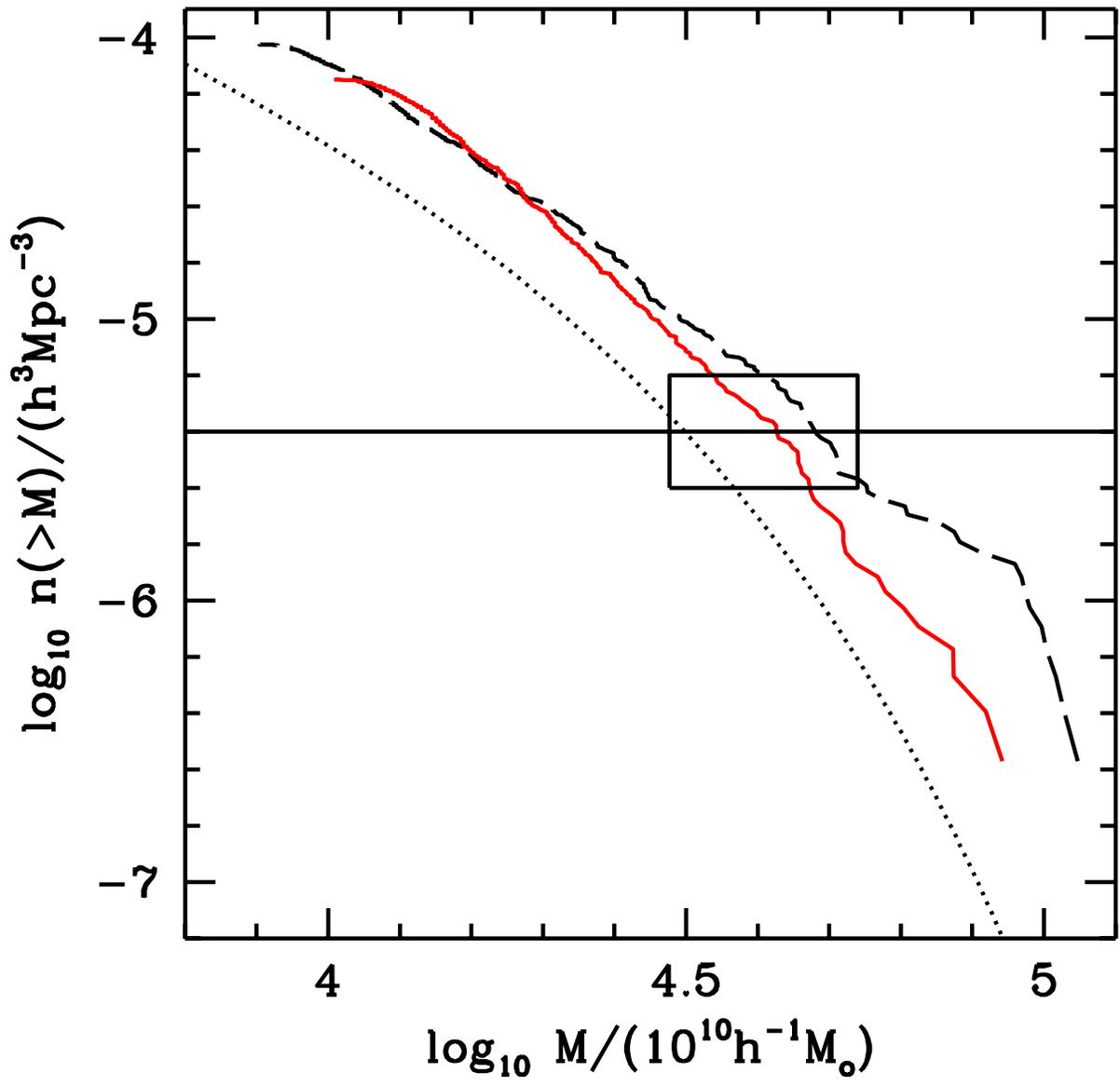
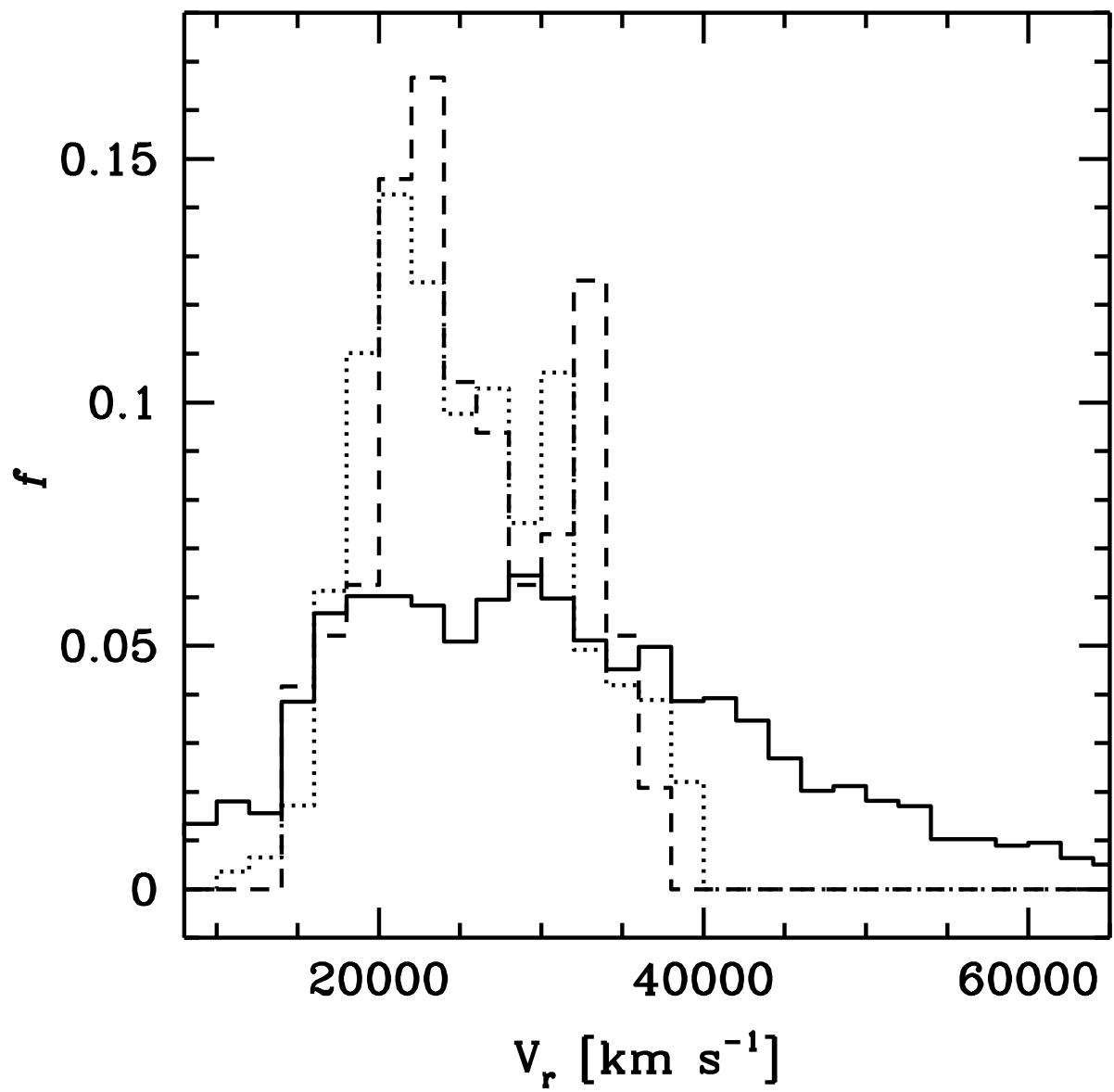
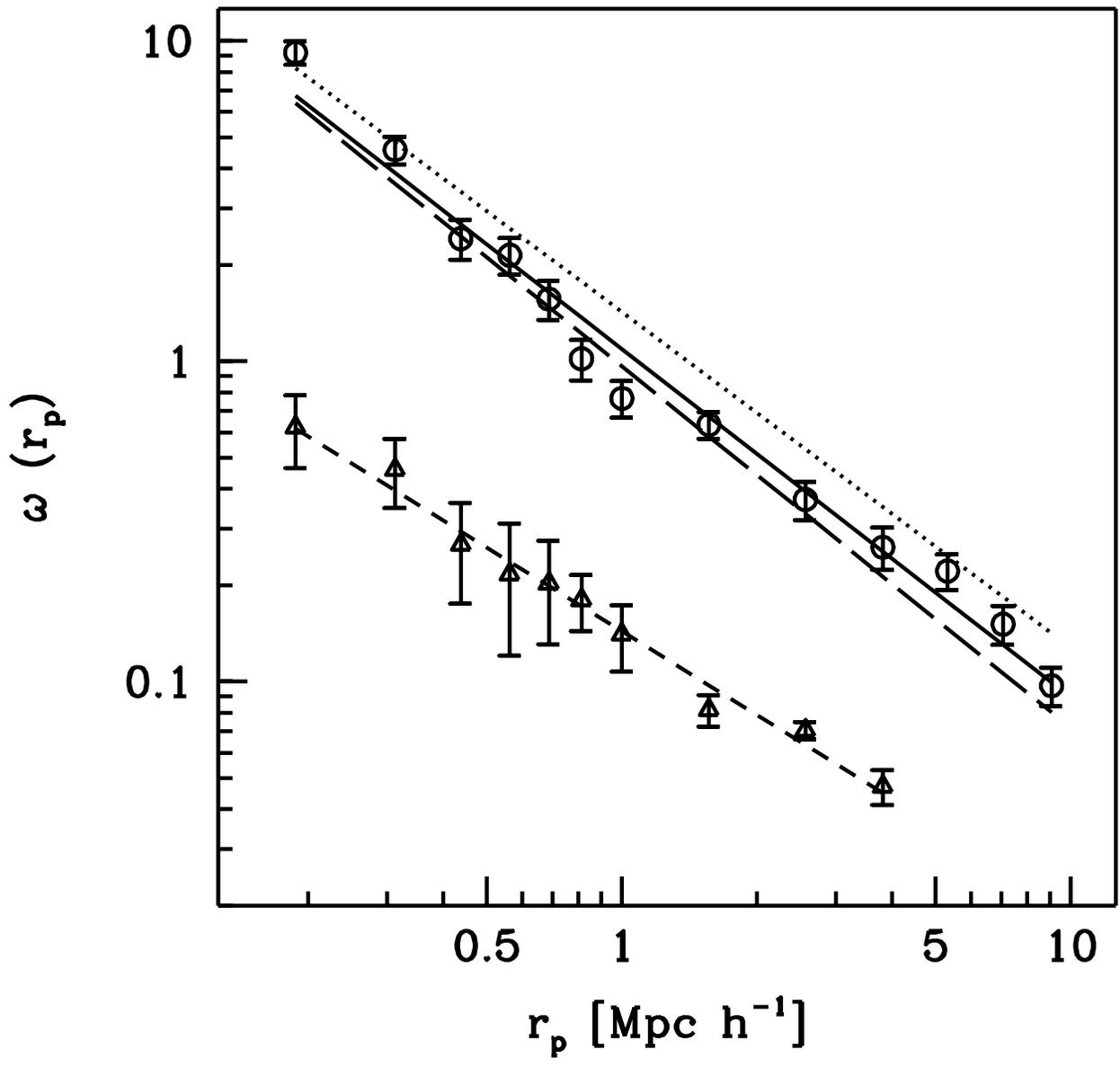
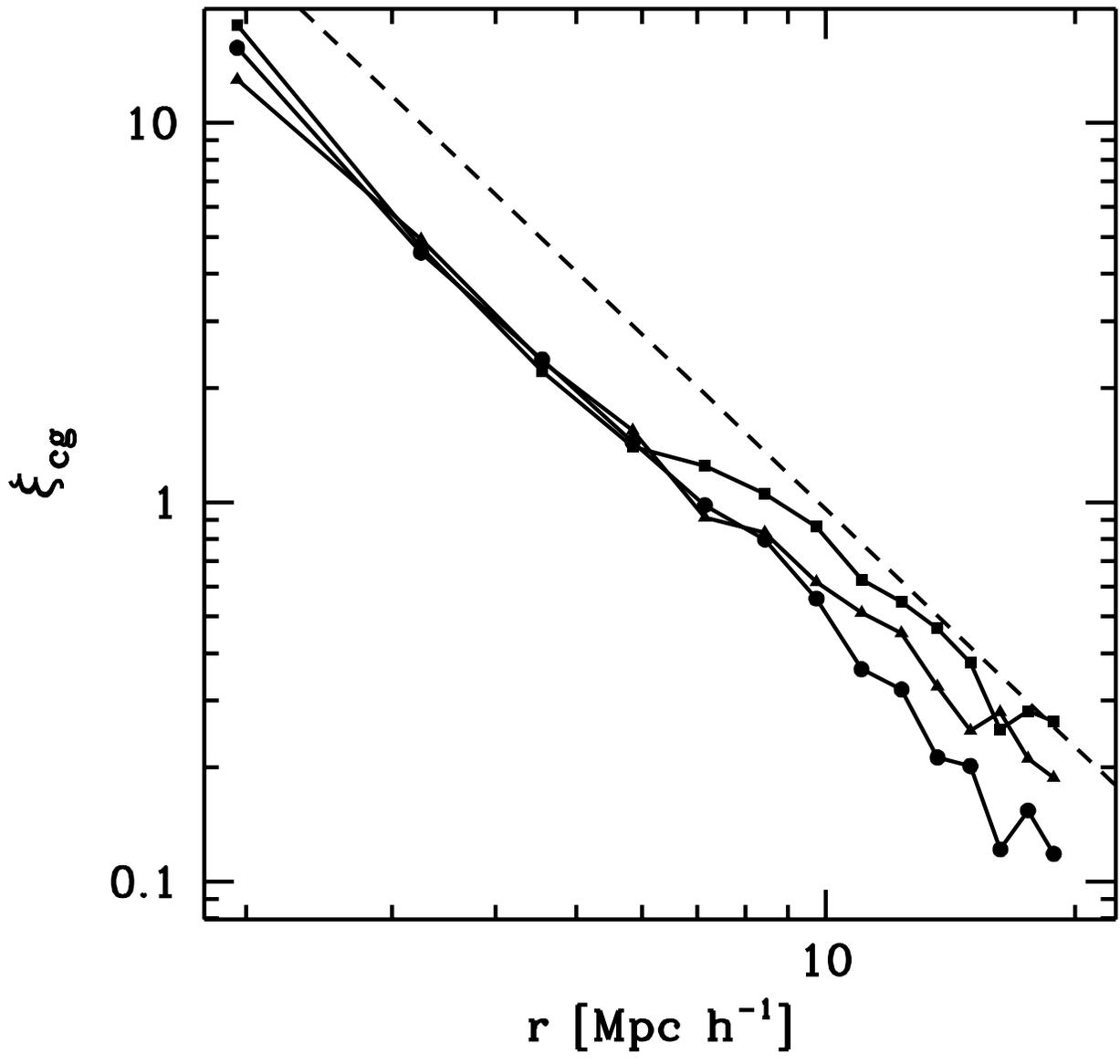
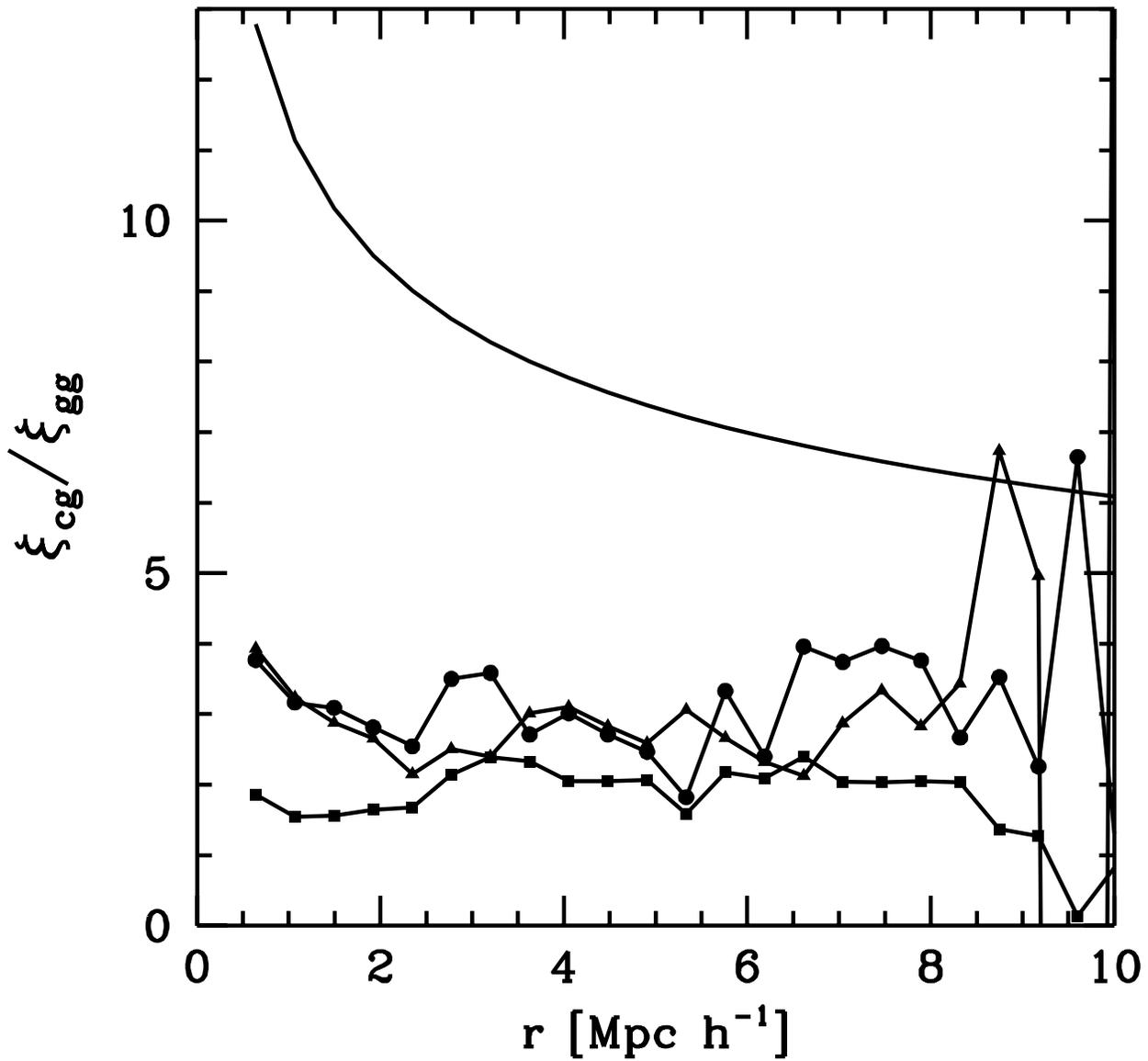
\end{document}